\documentclass[12pt, preprint]{aastex}
\usepackage{bbm}
\usepackage{natbib}
\usepackage{rotating}
\usepackage{amsmath}

\begin{document}

\title{Extragalactic Point Source Search in Five-year WMAP 41, 61 and 94~GHz Maps}

\author{X. Chen \& E. L. Wright}
\affil{Physics and Astronomy Department, University of California,
	Los Angeles, CA 90095-1547}

\begin{abstract}

We present the results of an extragalactic point source search using the five-year WMAP 41, 61 and 94 GHz (Q-, V- and W-band) temperature maps. This work is an extension of our designing and applying a CMB-free technique to extract point sources in the WMAP maps. Specifically, we have formed an internal linear combination (ILC) map of the three-band maps, with the weights chosen to remove the CMB anisotropy signal as well as to favor the selection of flat-spectrum sources. We have also constructed a filter to recover the true point source flux distribution on the sky. A total of 381 sources are found in our study at the $> 5\sigma$ level outside the WMAP point source detection mask, among which 89 are ``new'' (i.e., not present in the WMAP catalogs). Source fluxes have been calculated and corrected for the Eddington bias. We have solidly identified 367 ($96.3\%$) of our sources, the 1$\sigma$ positional uncertainty of which is 2$^\prime$. The 14 unidentified sources could be either extended radio structure or obscured by Galactic emission. We have also applied the same detection approach to simulated maps, which yielded 364$\pm$21 detections on average. The recovered source distribution $N(>S)$ agrees well with the simulation input, which proves the reliability of this method. 

\end{abstract}


\section{Introduction}

The Wilkinson Microwave Anisotropy Probe (WMAP) is designed to advance observational cosmology by making precise measurements of the cosmic microwave background (CMB) anisotropy (\citealt{2003ApJ...583....1B}). The characterizing and removing of foreground emission play a vital role in the interpretation of CMB temperature maps. At small angular scales, extragalactic radio sources, especially those with flat spectra, constitute the most important foreground. Therefore, it is essential to have efficient source detection techniques developed and applied to the WMAP maps. Meanwhile, since WMAP provides the only all-sky millimeter survey, source extraction in its maps also opens a window to study extragalactic radio sources in the millimeter wavelength region. 

With every WMAP data release, a catalog is provided of the brightest point sources in the WMAP maps. The detection procedure used by the WMAP science team features a filtering of the integration-time-weighted maps with $b_{\ell}/(b_{\ell}^{2}C_{\ell}^{CMB}+C_{\ell}^{noise})$ in harmonic space and a search for $> 5\sigma$ peaks, where $b_{\ell}$ is the transfer function of the WMAP beam response (\citealt{2003ApJS..148...39P}, \citealt{2007ApJS..170..263J}, \citealt{2009ApJS..180..246H}), $C_{\ell}^{CMB}$ is the CMB angular power spectrum and $C_{\ell}^{noise}$ is the noise power. This procedure successfully generated 208, 323 and 390 point sources in the WMAP first year, three-year and five-year maps, respectively (\citealt{2003ApJS..148...97B}, \citealt{2007ApJS..170..288H}, \citealt{2008arXiv0803.0577W}; hereafter, WMAP1, WMAP3 and WMAP5 catalogs). However, because of the limited angular resolution of WMAP, positive CMB excursions can be confused with point sources. In addition, the $C_{\ell}^{CMB}$ term in the filter counts as a systematic noise that does not integrate down with observing time. This largely limits the increasing of the number of detections with increased sensitivity of the maps. To circumvent the CMB ``noise'', \citet{2008ApJ...681..747C} introduced a CMB-free technique, which involves forming internal linear combination (ILC) of multi-frequency maps to suppress the CMB signal. The number of sources N found by applying this technique to the WMAP V- and W-band maps alone varied as  $t^{0.72}$ from one year to five years, in comparison to $N \sim t^{0.39}$ between the WMAP1 and WMAP5 catalogs (\citealt{2009ApJS..180..283W}).   

In this paper, we extend this CMB-free technique and employ it to the WMAP five-year Q-, V- and W-band temperature maps. The main goal is to utilize the high signal-to-noise ratio Q-band data to find more flat-spectrum sources, which are known to dominate the 30 - 100 GHz frequency regime.  
 
\section{Methodology}
\subsection{Point Source Detection}
The WMAP temperature maps at HEALPix\footnote{The Hierarchical Equal Area isoLatitude Pixelization (HEALPix) of the sphere is used to define WMAP map pixels on the sky in Galactic coordinates. Res 9 corresponds to 3,145,728 pixels over the full sky with a pixel resolution of $0.115\degr$. See more of HEALPix at http://healpix.jpl.nasa.gov/ . } Res 9 are used in this study. To minimize the noise difference between maps, we first smooth the high resolution V- and W-band maps to match the Q-band resolution. We construct the smoothing functions using a polynomial of an infinitely smooth function with compact support, i.e.,
\begin{equation}
 S(\theta)~=~\sum_n~a_nS_n~(\theta), 
\end{equation} where
\begin{equation}
S_n~(\theta) = \left\{
\begin{array}{ll}
\exp(- \frac {n \theta^2}{\theta_{Rc}^{2} - \theta^2}) & \theta < \theta_{Rc} \\
0 & \theta > \theta_{Rc}.
\end{array}\right.
\label{eq:smooth}
\end{equation}
The coefficients $a_{n}$'s are fitted by requiring the Legendre transform of the smoothing functions to match the ratios of Q-band beam transfer function to V-band and W-band beam transfer functions separately. And a cutoff radius $\theta_{R_{c}}$ of 1.25$\degr$~is chosen for the solid angle integrations. 

\begin{figure}
\plotone{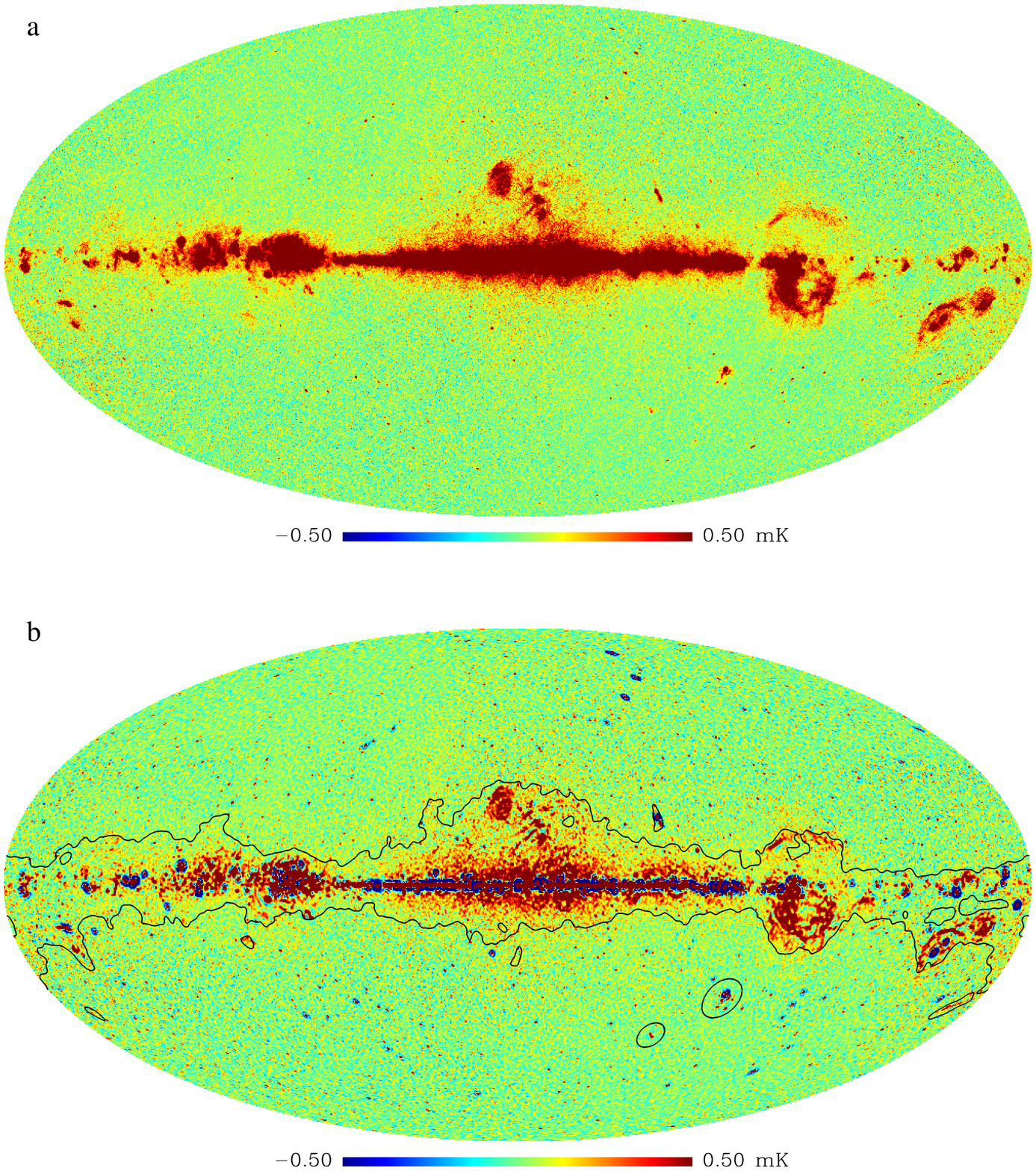}
\caption{(a) The ILC map constructed from the Q-band map, V-band map and W-band map, with the latter two smoothed to Q-band resolution. (b) The filtered ILC map with contours showing the Kp0+LMC+SMC mask used to exclude extended foreground emission. \label{ILC}}
\end{figure}

We then form the ILC map out of the Q-band map, V-band smoothed map and W-band smoothed map, 
\begin{equation}
T^{ILC} ~=~ \omega_Q T_Q + \omega_V T_V^{SM} + \omega_W T_W^{SM}.
\end{equation} 
The weights are determined by canceling out the CMB anisotropy signal
\begin{equation}
\omega_Q + \omega_V + \omega_W = 0, 
\end{equation}
normalizing the flat-spectrum source flux to the Q-band
\begin{equation}
\omega_Q + \omega_V (\frac {\partial B_Q}{\partial T} / \frac{\partial B_V}{\partial T})|_{T_0} + \omega_W (\frac {\partial B_Q}{\partial T} / \frac{\partial B_W}{\partial T})|_{T_0} = 1, 
\end{equation}
and minimizing the variance of the combination map, which can be approached by minimizing ($\omega_Q^2 + \omega_V^2 + \omega_W^2$) since the variance at each pixel is approximately the same for Q, V and W bands.
Here $B_{\nu}$ is the Planck function and $T_{0}$ = 2.725$\pm$0.002~K is the CMB temperature (\citealt{1999ApJ...512..511M}). The three-band ILC map is shown in Figure \ref{ILC}a.

For any pixel~$i$~within the Q-band beam to a point source, assuming negligible contribution from the overlap of point sources, its temperature $T_i^{ILC}$ in the ILC map can be fitted into the point source intensity $f$, multiplied by the beam response at that pixel location $b^Q_i$, plus a local baseline $c$, i.e., 
\begin{equation}
T_i^{ILC} ~=~ f b^Q_i ~+~ c.
\label{eq:srcfit}
\end{equation} 
Deviation from the fit is evaluated using the $\chi^2$, 
\begin{equation}
  \chi^2 ~=~ \sum_{ij} \epsilon_i ( N^{-1}_{ij} ) \epsilon_j,
\end{equation} 
where
\begin{equation}
  \epsilon_i ~=~ T_i^{ILC} ~-~ (f b^Q_i + c)
\end{equation} 
and
\begin{equation}
 N_{ij} ~=~\delta_{ij} \sigma_Q^2 \omega_Q^2 ~+~ \sum_k S^{V2Q}_{ik} S^{V2Q}_{jk} \sigma_V^2 \omega_V^2 ~+~ \sum_k S^{W2Q}_{ik} S^{W2Q}_{jk} \sigma_W^2 \omega_W^2.
\end{equation} 
Here $S^{V2Q}$ and  $S^{W2Q}$ are the smoothing functions used to smooth the V-band and W-band maps previously; the summation over $k$ is a sum over all the pixels within $\theta_{R_{c}}$ of pixel $i$ and $j$; $\sigma_Q$,  $\sigma_V$ and $\sigma_W$ are the noise at pixel $k$ in the Q-, V- and W-band maps, respectively. The best fit is defined when the $\chi^2$ reaches a minimum, which gives
\begin{equation}
\sum_{i j} N^{-1}_{i j} \left(
\begin{array}{c}
b_i^Q \\
1
\end{array}     \right) \left(
\begin{array}{cc}
b_j^Q & 1 \end{array} \right)  \left(
\begin{array}{c}
f \\
c
\end{array}     \right) = 
\sum_{i j} N^{-1}_{i j} T_j^{ILC} \left(
\begin{array}{c}
b_i^Q \\
1
\end{array}     \right).  
\end{equation}
Solving the above equation, we find that the intensity of the point source can be written as a weighted sum of the intensities of its surrounding pixels, i.e.,
\begin{equation}
f = \sum_{j} \left[\sum_{i} (M^{-1})_{00} b_i^Q N^{-1}_{i j} + \sum_{i} (M^{-1})_{01} N^{-1}_{i j}
\right] T_j^{ILC},
\label{eq:kernel}
\end{equation} 
where
\begin{equation}
M  =  
\sum_{i j} N^{-1}_{i j} \left(
\begin{array}{c}
b_i^Q \\
1
\end{array}     \right) \left(
\begin{array}{cc}
b_j^Q & 1 \end{array} \right). 
\end{equation}
Since the weights inside the square brackets of Equation (\ref{eq:kernel}) are essentially a function of the angular distance to the considered source, this suggests that we can probe the point source flux in the ILC map by fitting the function and filter the map locally. As a first attempt, we generate 32 random source positions in the sky and for each position calculate the weights to a distance of 1.25$\degr$. We plot the weights against the angular distances at each selected random spot, and note that they all share the same profile. This leads us to fit a single function using the collective weights from the 32 random positions. Convolving the all-sky ILC map with this global filter, we are able to recover the point source flux distribution on the entire sky, shown in Figure \ref{ILC}b. The extreme negative regions in the map are regions surrounding the brightest sources, where fluxes in the pixels are depleted and restored into their central point sources by filtering.

\begin{figure}
\plotone{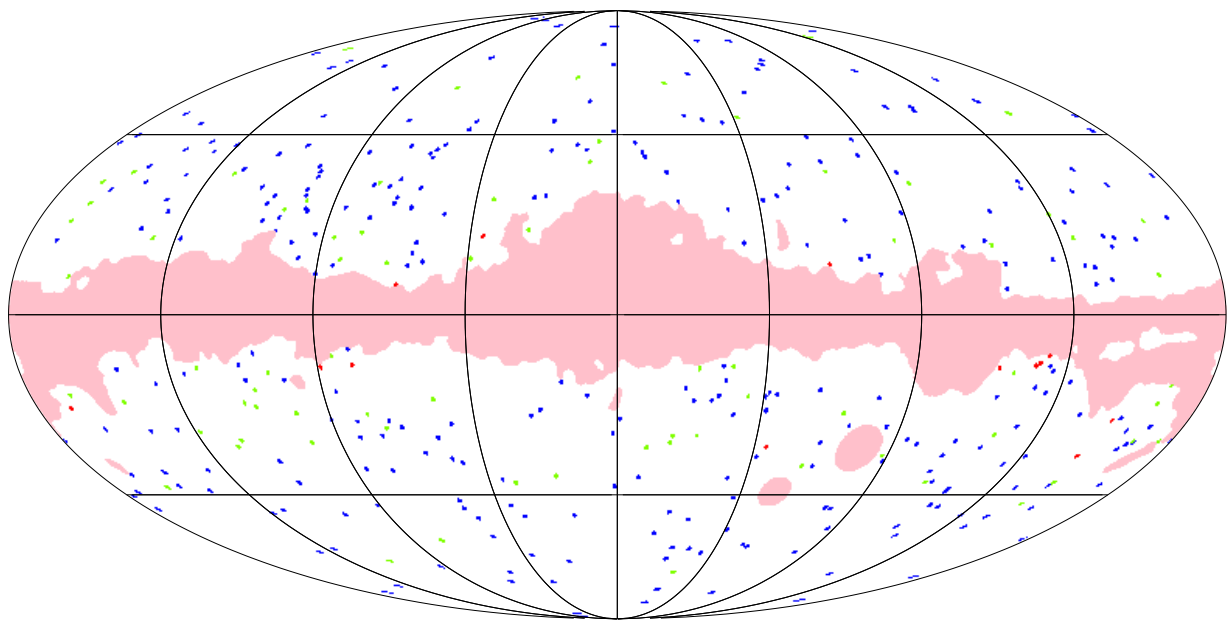}
\caption{Overview map showing the 381 point sources detected in the three-band ILC map. Blue dots represent sources in the WMAP catalogs, green indicates newly detected and identified sources, while red denotes the 14 unidentified sources. The shaded region shows the WMAP point source detection mask. \label{overview}}
\end{figure}

We adopt the same Kp0+LMC+SMC detection mask used by the WMAP team (\citealt{2009ApJS..180..283W}, shown as contours in Figure \ref{ILC}b) to exclude extended foreground emission. Following the WMAP scan pattern that gives non-uniform observation numbers among pixels:~greatest at the ecliptic poles, high at latitude $\pm45\degr$, and least in the ecliptic plane, we divide the masked sky into rings that cover 1 degree in ecliptic latitude. We then compute $\sigma$ individually within each ring and look for peaks greater than $5\sigma$. When more than one $> 5\sigma$ pixel lay within a 3 $\times$ 3 pixels area, the brightest one is chosen as a source detection. The accurate source position is approached by fitting a point source profile to the 9-pixel grid centered on the brightest pixel and defined at where the peak is. We find 381 sources in our study. The distribution of these sources on the sky is given in Figure \ref{overview}; Blue, green and red dots indicate, respectively, sources in the WMAP catalogs, newly detected and identified, and newly detected but unidentified (See \S \ref{id} for a detailed discussion of source identification). 

\subsection{Flux Estimation}
We estimate the Q-, V- and W-band flux densities of our sources by calculating
\begin{equation}
 F_{\nu} = \int I_{\nu}~\psi(\theta) \cos \theta d\Omega~=~\frac {\partial B_{\nu}}{\partial T} |_{T_{0}} \int T ~\psi(\theta) 
\cos \theta d\Omega,
\end{equation} 
where $B_{\nu}$ is again the Planck function and $T_{0}$ is the CMB temperature of 2.725$\pm$0.002~K. $T$ is the temperature measurement in each band;~the solid angle is integrated to a radius of 1.25 degree. We have introduced a weighting function $\psi(\theta)$ in our flux estimator to help enhance the contrast of the point source flux to the background. This function is constructed individually in each band with the corresponding beam profile in the center and a negative Gaussian ring in the outer field. We require
\begin{equation}
   \int \psi(\theta) d\Omega = 0,
\end{equation}
in order to ignore any flux that spreads uniformly across the whole integration field. The mean and variance of the Gaussian functions are optimized to preserve the source flux while minimizing the fluctuation in the integration field: 
\begin{equation}
\sigma^2 = \sum_i \sum_j \psi_i \psi_j (C(\theta_{ij}) + \delta_{ij} N_i), 
\end{equation} 
where
\begin{equation}
C(\theta) = \sum_{\ell} \frac {2\ell + 1}{4 \pi} C_{\ell} \omega_{\ell} P_{\ell} (cos \theta).
\end{equation}
Here i and j denote pixels within the integration field; N is the radiometer noise in the five-year maps; $C_{\ell}$ is the best fit power spectrum of WMAP5 data assuming a $\Lambda$CDM model; $\omega_{\ell}$ is the window function which encodes the beam smoothing (\citealt{2003ApJS..148...39P}); $P_{\ell}$ is the Legendre function. We see negative fluxes for some sources, which is likely caused by our source sitting on the top of a negative CMB fluctuation. The error on the flux density is calculated as the flux rms value of the background pixels that are close to the target source but are not ``contaminated'' by it. Specifically, we select a ring of pixels around the source with an inner radius equal to 1$\degr$ to exclude pixels affected by the source fluxes, and an outer radius of 2$\degr$ to include sufficient number of pixels to estimate the background fluctuation level. 

\begin{figure}
\epsscale{0.75}
\plotone{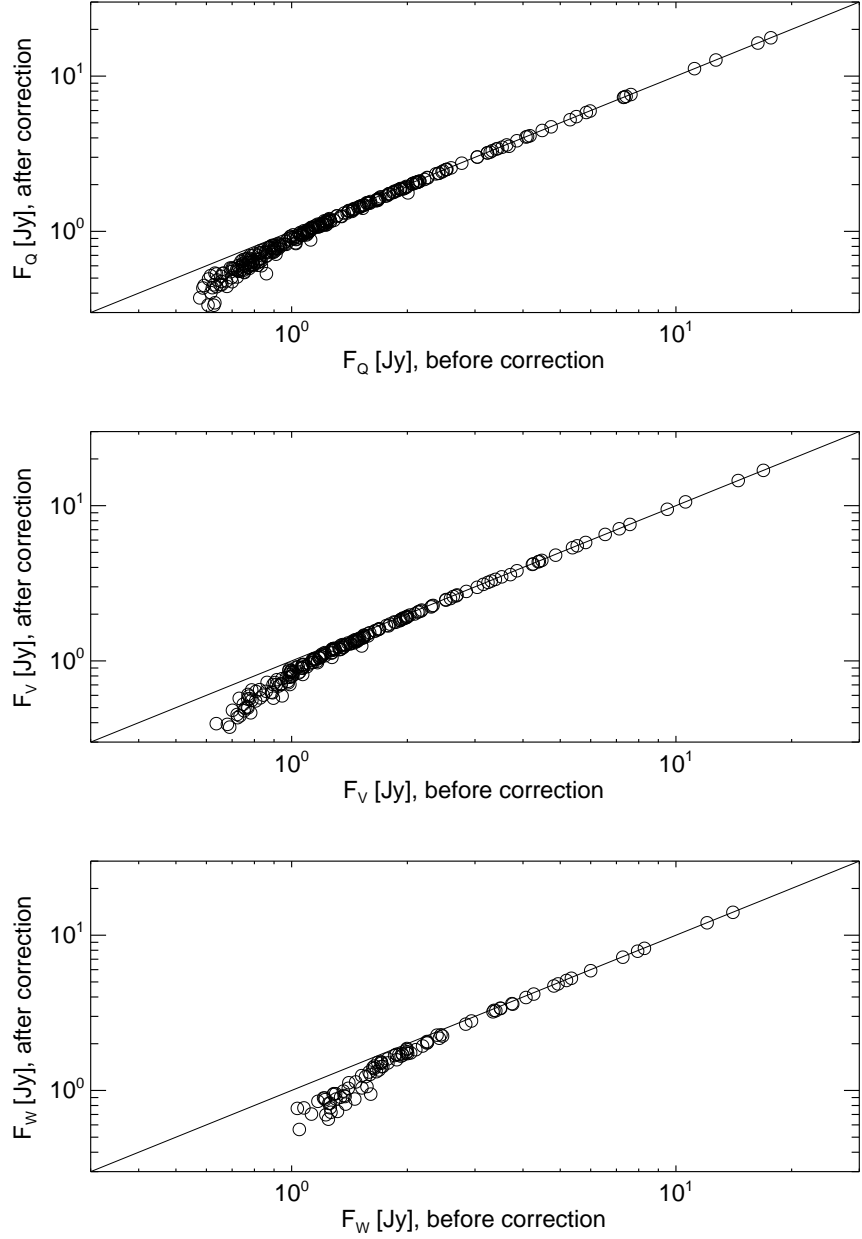}
\caption{Comparison of our flux estimates before and after Bayesian correction. \label{eddington}}
\end{figure}

Since flux estimates of sources in noisy fields are on average overestimated (Eddington Bias, \citealt{1940MNRAS.100..354E}), we have applied the Bayesian approach described in the Appendix B of \citet{2007ApJS..170..108L} to correct for the bias. The estimated slopes of the differential number counts are 2.67, 2.68 and 2.30 for the Q, V, and W bands separately. We compare the uncorrected fluxes of each band with the corrected ones in Figure \ref{eddington}. The agreement is generally good, and the correction makes almost no difference for fluxes $\ga$ 1.5 Jy in Q and V bands.   

\subsection{Identification}\label{id}

\begin{figure}
\epsscale{0.8}
\plotone{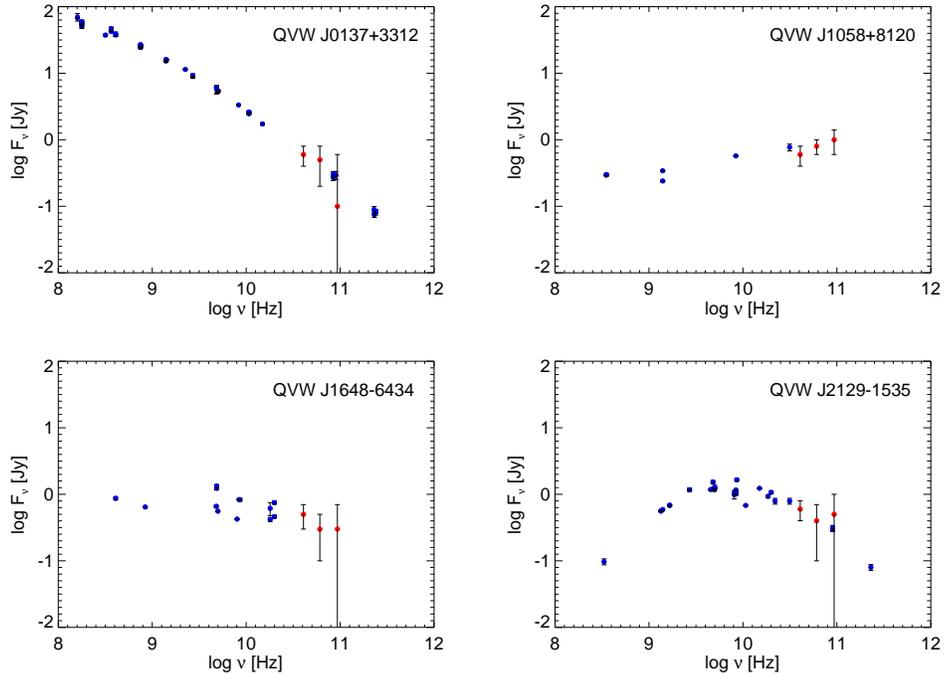}
\caption{Radio spectra examples of our sources. Red and blue dots indicate, respectively, flux estimates obtained in this work and from NED. \label{sedexample}}
\end{figure}

\begin{figure}
\epsscale{0.75}
\plotone{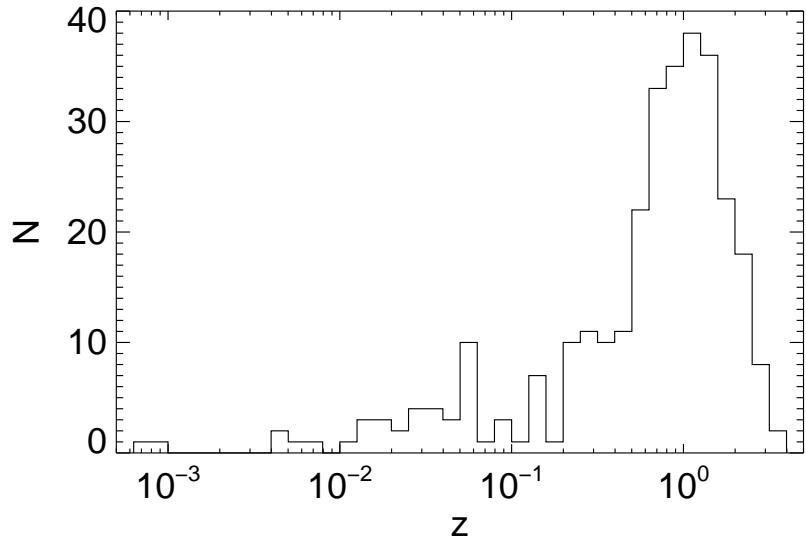}
\caption{Redshift distribution of the 306 sources that have redshift values given in NED. The median of z is 0.85. \label{redshift}}
\end{figure}

We search in the NASA/IPAC Extragalactic Database (NED\footnote{See \url{http://nedwww.ipac.caltech.edu/}.}) and the source catalogs from two recent all/large sky radio surveys: CRATES~(8.4 GHz, \citealt{2007ApJS..171...61H}) and the Australia Telescope 20 GHz Survey (AT20G, \citealt{2008MNRAS.384..775M}), for radio sources within 15$^\prime$ ($\theta^{Q}_{FWHM} \sim 30.6^\prime$) to our sources. We retrieve the available photometric data and make the spectral energy distribution (SED) plots for these radio sources. If the flux densities of our source can be reasonably fitted into the SED of a known radio source, we associate them. This SED approach identifies 367 (i.e. 96.3$\%$) sources on our list, the majority of which are found to have a flat spectrum, i.e., ${\alpha} \sim 0 $ in $F_{\nu} \sim \nu^{\alpha}$. This is consistent with our knowledge of the dominance of flat-spectrum sources in this wavelength regime, and with the weights we chosen for the ILC map that favors flat-spectrum sources. Nevertheless, we do detect some steep-spectrum sources as well as sources with spectra peaked at $\sim$10 GHz (see Figure \ref{sedexample} for some SED examples). We have also searched NED for the optical identifications of our sources and found 269 QSOs, 61 galaxies, 1 galaxy pair, 1 galaxy triple, 1 planetary nebula and 34 unclassified radio sources among the identified ones. Additionally, NED provides redshifts for 306 identified sources. A plot of the redshift distribution of these sources is given in Figure \ref{redshift}, the median of z is 0.85. 

\begin{figure}
\plotone{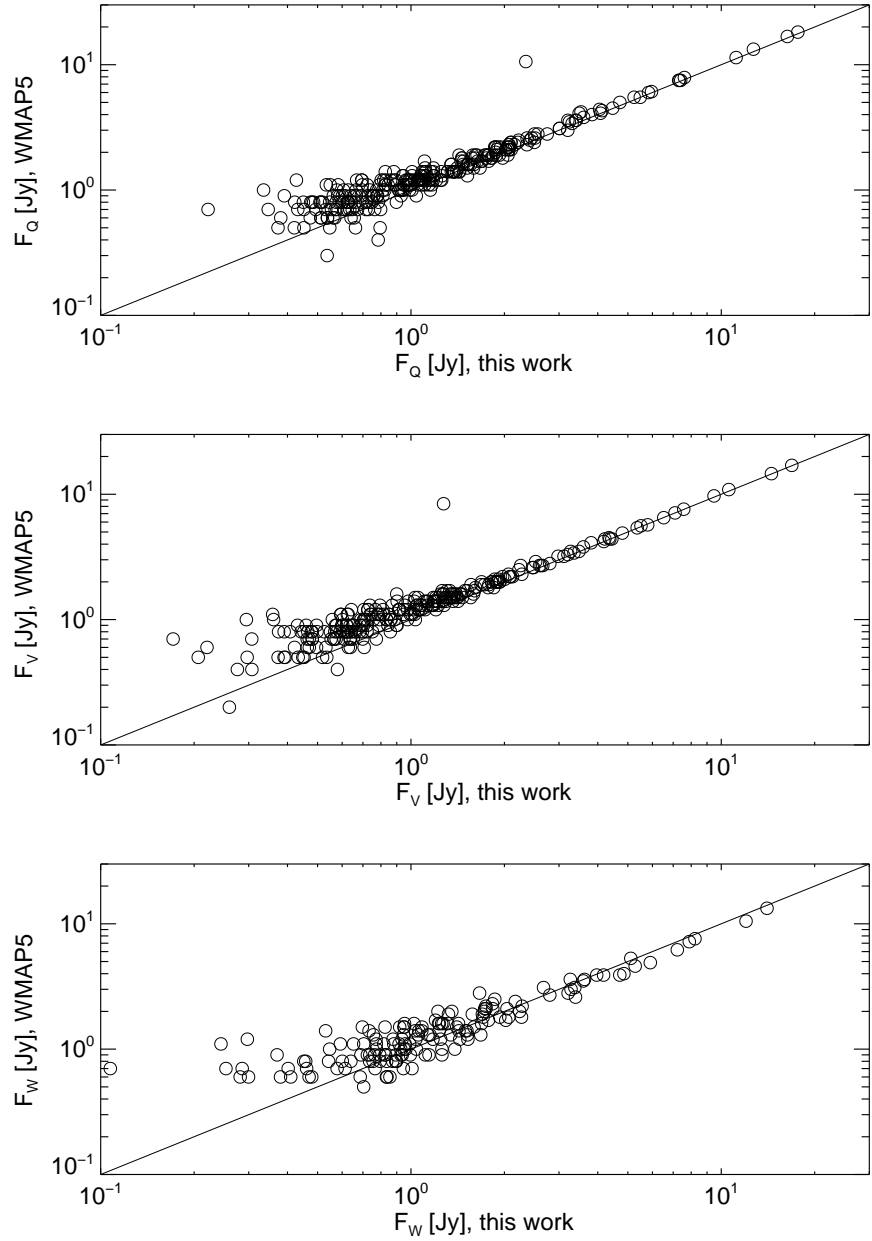}
\caption{Comparison of our flux estimates with WMAP5 flux densities at Q, V and W bands. \label{fluxcompare}}
\end{figure}

We then cross-correlate our list of identified sources with the WMAP catalogs. If a source is within 15$^\prime$ of a WMAP cataloged source, we tag it. We find 287 of our sources present in the WMAP5 catalog, and five more sources (J0734+5026, J1513-1013, J1553-7914, J1644-7713, J1648+4109) are in the WMAP3 catalog but missing from the WMAP5 catalog. The high associate rate clearly demonstrates that our method can generate results consistent with the method adopted by the WMAP team. In Figure \ref{fluxcompare}, we compare our flux estimates with the ones given in the WMAP catalogs. These two sets of fluxes agree very well with each other at $F_{\nu} \ga$ 1 Jy. Our estimates are in general lower than the WMAP ones for faint sources, which is partially due to the flux correction we made. The most deviated point in both Q and V bands is Fornax A, known as an extended source with a relatively weak core and two large extended lobes. A complete list of all the identified sources is given in Table \ref{src_id},  sorted in ascending order of right ascension. For each source, we give its WMAP5 ID (if available), optical identification, flux densities along with 1$\sigma$ errors at the WMAP Q-, V- and W-band, its 5 GHz ID as given in GB6~(\citealt{1996ApJS..103..427G}), PMN~(\citealt{1994ApJS...90..179G, 1995ApJS...97..347G}, \citealt{1994ApJS...91..111W, 1996ApJS..103..145W}),  S5 (\citealt{1981AJ.....86..854K}) or \citet{1981A&AS...45..367K} catalogs,  and the angular distance to its 5 GHz counterpart. For four of our sources (J0422+0212, J0721+0403, J0805+6143, J1015+2258), we have assigned different 5GHz IDs than given in the WMAP5 catalog based on the SEDs of the 5 GHz sources. The position errors of our detections are evaluated with respect to the positions of their 5 GHz counterparts since the 5 GHz surveys have in general higher angular resolutions. Assuming the source deviations from their 5 GHz counterparts in both the galactic longitude and latitude axes are normally distributed, the angular distance $r$ to the 5 GHz IDs should satisfy a Rayleigh distribution and the radial positional uncertainty of our sources can then be estimated as $\sigma_r = median(r)/ \sqrt {\ln 4} = 2^\prime$. We conclude that our 5 GHz IDs are more reliable since we obtain a better positional accuracy than the WMAP5 survey (4$^\prime$ in both longitudes and latitudes). 

There are 14 sources left without any solid identification, as listed in Table \ref{src_uid} and noted in red in Figure \ref{overview}. Looking back into the filtered ILC map (Figure \ref{ILC}b), we do see bright emission features at the corresponding locations. We suggest that these sources could be either extended radio structure or obscured by Galactic emission. Considering the low or negative flux densities obtained for some of these sources, it is likely that some of them are located on the top of negative CMB fluctuations so that they do not show as strong sources in each band but show up in the CMB-free ILC map. We cross-correlate this list with the GB6, PMN, S5 and \citet{1981A&AS...45..367K} catalogs. If a 5 GHz source is within 15$^\prime$, we suggest it as the possible counterpart of our source. In cases when no 5 GHz source is found nearby, we use the closest 1.4 GHz NVSS (\citealt{1998AJ....115.1693C}) source or 843 MHz SUMSS (\citealt{1999AJ....117.1578B}, \citealt{2003MNRAS.342.1117M}) source instead.

\section{Discussion}
\subsection{Source Number Counts}
\begin{figure}
\epsscale{0.75}
\plotone{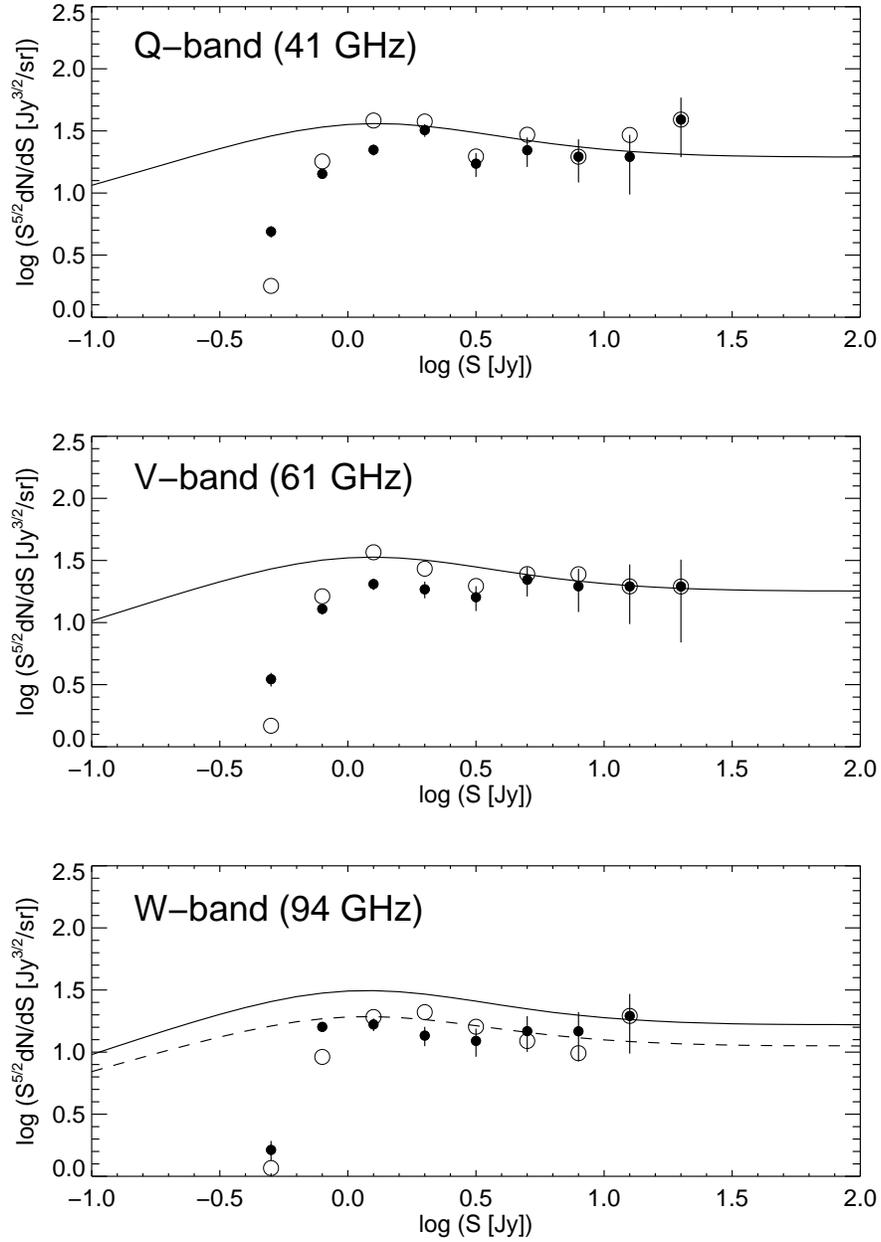}
\caption{Point source counts derived at the Q-, V- and W-band, in bins of $\Delta$log S = 0.2.  Sources detected in the ILC map are shown in filled circles with Poisson error bars. Sources in the WMAP5 catalog are shown in big circles for comparison. The curves show the counts predicted by the \citet{2005A&A...431..893D} model. The dashed line in the lowest panel shows the same model scaled by 0.86.  \label{numcnt}}
\end{figure}

We have calculated the source number counts in bins of $\Delta$log S = 0.2 at Q-, V- and W-band\footnote{By convention, S is often used to represent the flux density in the discussion of number counts distribution; it is equivalent to $F_{\nu}$ that is used throughout this paper. } and compared them with the predictions of the cosmological evolution model by \citet{2005A&A...431..893D}. As illustrated in Figure \ref{numcnt}, the agreement is generally good above $\sim$ 2 Jy (i.e., log S = 0.3) in the Q and V bands. The low data points at lower fluxes are due to the incompleteness of our sources at these flux ranges. In the W-band, our data points agree more with a model rescaled by 0.86. The source counts derived from WMAP5 catalog in the corresponding bands are also plotted for comparison. They are mostly consistent with the source counts in this work.  

\subsection{Chance-Coincidence Rate}
\begin{figure}
\plotone{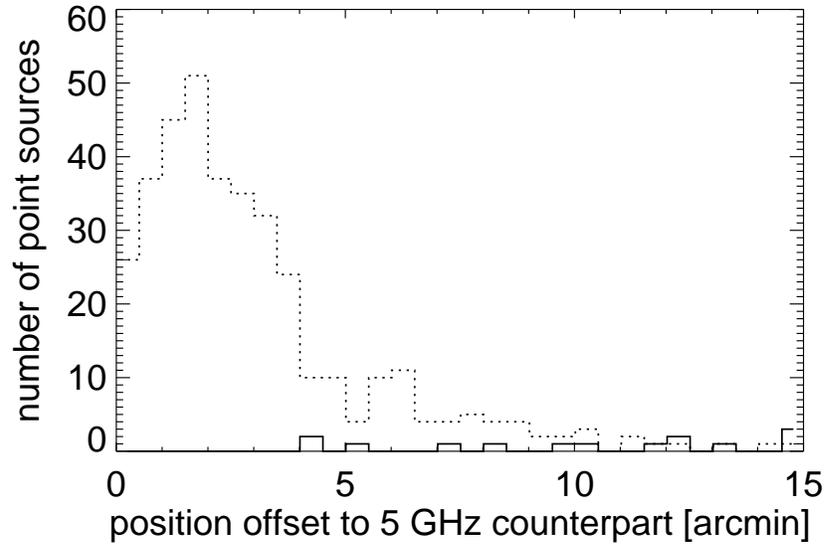}
\caption{Distribution of the deviations from the real detections (dotted line) and false identifications with random positions (solid line) to their 5 GHz counterparts. \label{posierr}}
\end{figure}

To estimate the influence of false identifications from random coincidences, we have generated an equal number (i.e., 381) of random positions outside the WMAP source detection mask. We repeat the same near-position search in NED. If a 5 GHz source is found to have $F_{\nu} \ga$ 100 mJy within 15$^\prime$ of a fake source position, we calculate the spectral index $\alpha$ using the 5 GHz flux estimate along with the 1.4 GHz flux estimate from the NVSS catalog or 843 MHz flux estimate from the SUMSS catalog (all such 5 GHz sources have a NVSS or SUMSS counterpart). We then use the spectral index to extrapolate this source's 5 GHz flux to the WMAP 41 GHz channel (Q-band). We consider a source as identified when its Q-band flux is above 100 mJy. Following this approach, we associate 14 random sources with known radio sources, corresponding to a chance-coincidence rate of 3.7$\%$. In Figure \ref{posierr}, we plot the distribution of position offsets of real detections to their 5 GHz counterparts, in comparison with the distribution of the identified random sources. It is evident that the false identifications with random sources generally have a larger position error than the real peak positions. Based on this plot, we suggest that our source identifications within 5$^\prime$ are probably real. 

\subsection{Analysis of Simulated Maps}
\begin{figure}
\plotone{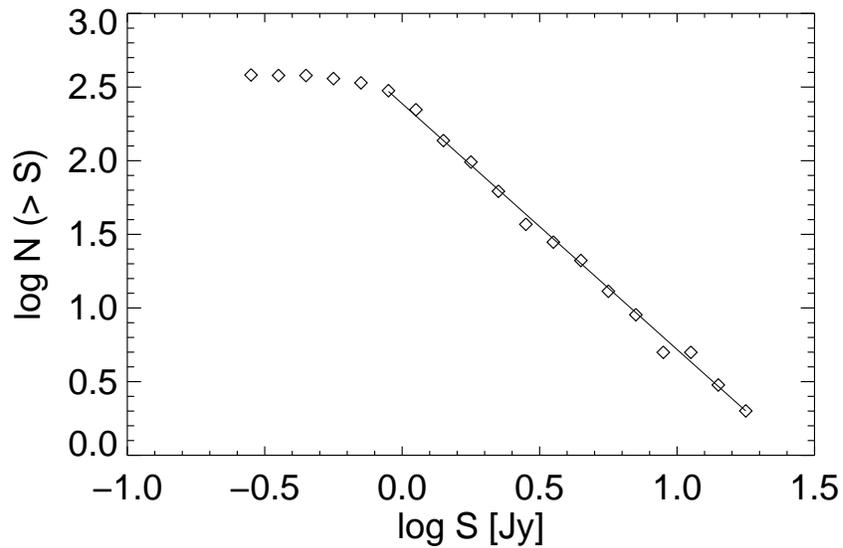}
\caption{Integral counts of extragalactic sources in the WMAP5 Q-band, centered at 40.7 GHz.\label{intcnt}}
\end{figure}

We repeat the same point source analysis on simulated maps constructed with point sources, CMB fluctuations, and radiometer noise. $10^6$ sources are sampled from a power law distribution $N(>S)$ at the WMAP Q-band (centered at 40.7 GHz), derived from WMAP5 sources with $F_Q > 1$ Jy as they appear to be complete to $\sim 1$ Jy (Figure \ref{intcnt}). Spectral indices are chosen from a Gaussian with mean -0.09 and standard deviation 0.176 (\citealt{2009ApJS..180..283W}), and the fluxes are scaled to the V- and W-band centers (60.8 GHz and 93.5 GHz). In each band, we calculate the appropriate temperature of every source and assign it to a random HEALPix pixel at Res 11 (a total of $12\times4^{11}$ pixels). These point source maps are then smoothed with the beam window function at each band and converted to Res 9 maps. Finally, we add in the Res 9 maps of CMB fluctuations produced using the best fit $C_{\ell}^\prime$s from WMAP5 data assuming a $\Lambda$CDM model, and of radiometer noise generated from the WMAP5 noise variance in each pixel. The point source detection process is then applied to these simulated maps, yielding on average 364$\pm$21 point sources for the 10 simulations we completed. When compared with the input point source maps, we find only 2 spurious sources out of the total detections from our simulations. The recovered $N(>S)$ agree remarkably well with the simulation input for fluxes $> 1$~Jy, which infers that our method is robust and the source counts we get from the WMAP5 ILC map are reliable for fluxes $> 1$~Jy.

\section{Conclusions}
We present catalogs of point sources found in an ILC map constructed from the WMAP five-year Q-, V- and W-band maps. We have recovered 287 WMAP5 sources and five WMAP3 sources in our survey. Bypassing of CMB ``noise'' in the detection process brings us 89 new sources (23.4$\%$ of the total). We have obtained flux density estimates for our sources in each of the three bands. Most of the new sources are found to have low or even negative flux estimates in at least one band, which is probably due to the coincidence of their locations with negative CMB fluctuations. This will generally lead to missing of these sources in single frequency band searches. We identify our sources by searching in NED for nearby low frequency radio sources and fitting the fluxes of our sources to their SEDs. Of all the sources we detected, we lack of solid identifications for 14. The flux estimates and number count distribution of our identified sources are comparable with those from the WMAP5 catalog, which proves our method to be complimentary to as well as consistent with the WMAP approach. Since this CMB-free technique responds rapidly as observing time increases (\citealt{2009ApJS..180..283W}), we expect a fast increase of new detections in more years of WMAP maps.

\acknowledgments
We acknowledge the use of the Legacy Archive for Microwave Background Data Analysis (LAMBDA) to retrieve the WMAP data set. 
Support for LAMBDA is provided by the NASA Office of Space Science. This research has also made use of the VizieR catalog service and the HEALPix (\citealt{2005ApJ...622..759G}) package.

\clearpage

\begin{deluxetable}{lrlcrrrccc}
\tablecolumns{10}
\tabletypesize{\scriptsize}
\tablewidth{0pt}
\tablecaption{Point Sources from 5-year Q-, V-, and W-band ILC Map - Identified \label{src_id}} 						
\tablehead{
\colhead{RA} & \colhead{DEC}  &
\colhead{WMAP ID} & \colhead{Type} & \colhead{Q} & \colhead{V} & \colhead{W}& \colhead{5 GHz ID} & \colhead{Dist.} & \colhead{Note }\\
\colhead{hms} & \colhead{dm} & \colhead{} &
\colhead{}  & \colhead{[Jy]} & \colhead{[Jy]}  & \colhead{[Jy]} &  \colhead{} & \colhead{[arcmin]} & \colhead{} 
}
\startdata 
\hline
  00 06 13 & -06 24 & J0006-0623  & G       &   2.4$\pm$0.2 &   1.8$\pm$0.3 &   1.0$\pm$0.5 & PMN J0006-0623  &  0.9   &  b    \\
  00 10 34 & 10 58  & J0010+1101  & G      &   1.0$\pm$0.2 &   0.7$\pm$0.2 &   1.3$\pm$0.4 & GB6 J0010+1058   &  0.8      &   b      \\
  00 13 05 & -39 53 & J0012-3952  & RadioS  &   0.7$\pm$0.2 &   1.0$\pm$0.2 &   1.1$\pm$0.4 & PMN J0013-3954    &  1.1    &   b c    \\
  00 13 45 & 40 51  & \nodata          & G       &   0.5$\pm$0.2 &   0.6$\pm$0.2 &   0.1$\pm$0.4 & GB6 J0013+4051   &  2.6     & a b      \\
  00 19 40 & 26 00  & J0019+2603  & QSO     &   0.9$\pm$0.2 &   0.4$\pm$0.3 &   0.5$\pm$0.4 & GB6 J0019+2602  &  2.1      &   b      \\
  00 19 56 & 20 20  & J0019+2020  & RadioS  &   0.6$\pm$0.2 &   0.8$\pm$0.3 &  -0.4$\pm$0.4 & GB6 J0019+2021   &  4.6     &   b      \\
  00 26 04 & -35 13 & J0026-3510  & RadioS  &   1.4$\pm$0.2 &   1.0$\pm$0.2 &   0.9$\pm$0.4 & PMN J0026-3512    &  2.5    &  b c    \\
  00 37 56 & -02 07 & \nodata          & G       &   0.5$\pm$0.2 &   0.4$\pm$0.2 &   0.9$\pm$0.4 & PMN J0038-0207    &  6.1    &  \nodata       \\
\enddata
\tablenotetext{a}{The source position is fitted to the local maximum of a 5 $\times$ 5 pixel grid as there are two bright pixels within the Q-band beam, both of which are the brightest in their 3 $\times$ 3 pixel neighborhood.}
\tablenotetext{b}{Indicates this source is included in the CRATES catalog. }
\tablenotetext{c}{Indicates this source is included in the AT20G BSS catalog. }
\tablenotetext{d}{Indicates the source is in the three-year WMAP point source catalog, but not in the five-year catalog.}
\tablenotetext{e}{The source has a a different 5 GHz ID than noted in the WMAP five-year point source catalog.}
\end{deluxetable}

\begin{deluxetable}{lrrrrcccc}
\tablecolumns{9}
\tabletypesize{\scriptsize}
\tablewidth{0pt}
\tablecaption{Point Sources from 5-year Q-, V-, and W-band ILC Map - Unidentified\label{src_uid}} 						
\tablehead{
\colhead{RA} & \colhead{DEC}  &
\colhead{Q} & \colhead{V} & \colhead{W}& \colhead{Possible Radio Companion\tablenotemark{a}} & \colhead{Dist.} & \colhead{$f_{\nu}$\tablenotemark{b}} & \colhead{Note} \\
\colhead{hms} & \colhead{dm} &
\colhead{[Jy]} & \colhead{[Jy]}  & \colhead{[Jy]} &  \colhead{} & \colhead{[arcmin]}  & \colhead{[Jy]} & \colhead{} 
}
\startdata 
\hline
  03 58 37 & 22 36  &   0.4$\pm$0.2 &   0.4$\pm$0.3 &   0.9$\pm$0.5 & GB6 J0358+2236 & 0.4 & 0.074 & \nodata \\
  04 34 54 & -10 05 &   0.3$\pm$0.2 &  -0.2$\pm$0.2 &  -0.6$\pm$0.4 & PMN J0434-1007  & 4.2  & 0.121 &  c  \\
  05 03 45 & -04 10 &   0.1$\pm$0.2 &  -0.2$\pm$0.3 &   0.4$\pm$0.5 & NVSS 050342-041420  & 3.5  & 0.079 & d   \\
  06 43 44 & -24 11 &  -0.0$\pm$0.2 &  -0.1$\pm$0.2 &   0.2$\pm$0.4 & NVSS J064353-240440  & 7.4 & 0.018  &  \nodata  \\
  06 44 28 & -23 09 &   0.2$\pm$0.2 &   0.2$\pm$0.2 &   0.5$\pm$0.4 & PMN J0644-2322  &  13.7 & 0.043 &   \nodata   \\
  06 44 47 & -24 34 &   0.1$\pm$0.2 &  -0.2$\pm$0.2 &   0.2$\pm$0.3 & NVSS J064419-243804  & 7.2 & 0.004 &   \nodata \\
  06 44 49 & -23 31 &  -0.1$\pm$0.2 &  -0.2$\pm$0.2 &  -0.2$\pm$0.4 & PMN J0644-2322  & 8.9 & 0.043 & d    \\
  06 47 20 & -20 26 &   0.5$\pm$0.2 &  -0.1$\pm$0.2 &  -0.2$\pm$0.4 & NVSS J064748-201517 & 13.1 & 0.020 & \nodata \\
\enddata
\tablenotetext{a}{The 5 GHz radio source within 15 arcmin of our source position is given. When a 5 GHz ID is not available, the closest source from NVSS and SUMSS survey is given.}
\tablenotetext{b}{The corresponding flux density of the possible radio companion at 4.85 GHz for PMN and GB6 sources, 1.4 GHz for NVSS sources and 0.843 GHz for SUMSS sources.}
\tablenotetext{c}{This source is in the CRATES catalog, with an 8.4 GHz flux measured at 0.074 Jy.}
\tablenotetext{d}{There are two pixels within one Q-band beam, both of which are the brightest in their 3 by 3 pixel neighborhood. The location showing here is fitted to the local maximum of a 5 $\times$ 5 pixel neighborhood including both bright sources. }
\end{deluxetable}

\end{document}